\documentclass[journal=jacsat,manuscript=article]{achemso}
\setkeys{acs}{articletitle = true}
\setkeys{acs}{maxauthors = 99}
\usepackage[version=3]{mhchem} 
\usepackage[T1]{fontenc}       

\usepackage{subfig}
\usepackage{graphicx}
\usepackage{dcolumn}
\usepackage{bm}
\usepackage[flushleft]{threeparttable}
\usepackage{url}
\usepackage{amsmath}
\usepackage{amssymb}
\usepackage{braket}
\usepackage{multirow}
\usepackage{array}
\usepackage{url}
\usepackage{booktabs}
\usepackage{float}
\newcolumntype{P}[1]{>{\centering\arraybackslash}p{#1}}
\usepackage{color,soul}
\usepackage[usenames,dvipsnames]{xcolor}
\newcolumntype{L}{>{\centering\arraybackslash}m{3cm}}
\usepackage{lipsum} 

\author{Jing Yang}
\affiliation[University of Pennsylvania]
{Department of Chemistry, University of Pennsylvania, Philadelphia, Pennsylvania 19104--6323, USA}
\author{Yubo Qi}
\affiliation{Department of Chemistry, University of Pennsylvania, Philadelphia, Pennsylvania 19104--6323, USA}
\author{Hyeong-Seok D. Kim}
\affiliation{Department of Chemistry, University of Pennsylvania, Philadelphia, Pennsylvania 19104--6323, USA}
\author{Andrew M. Rappe}
\affiliation{Department of Chemistry, University of Pennsylvania, Philadelphia, Pennsylvania 19104--6323, USA}
\email{rappe@sas.upenn.edu}
\phone{+1 (215) 898-8313}
\fax{+1 (215) 573-2112}

\title{Tribopolymer Formation Mechanism on the RuO$_2$(110) Surface}


\begin{document}

\begin{tocentry}
\begin{figure}[H]
\begin{center}
\includegraphics[scale=0.1]{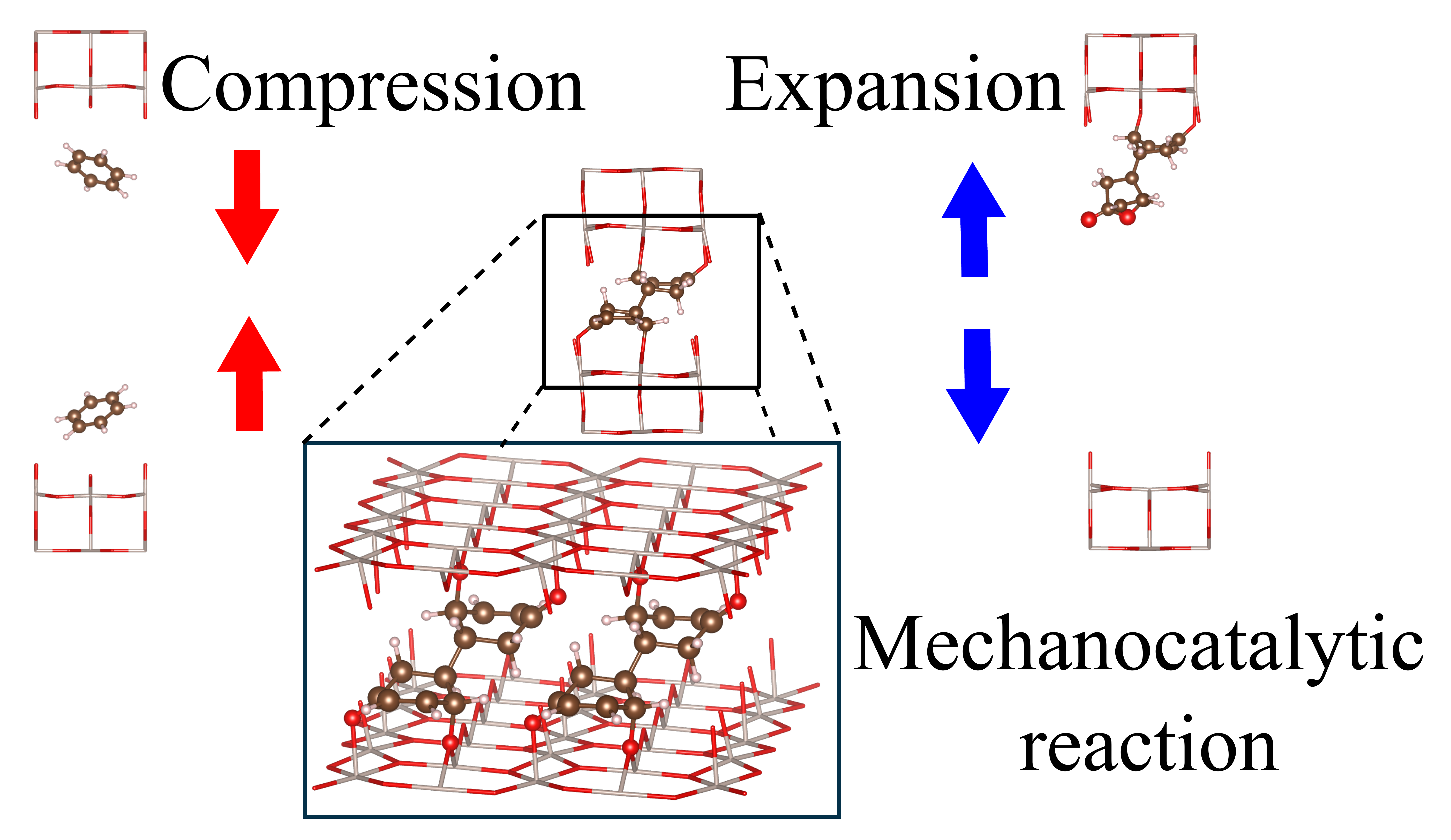}
\end{center}
\end{figure}
\end{tocentry}

\begin{abstract}
Tribopolymer formed on the contacts of microelectromechanical and nanoelectromechanical system (MEMS/NEMS) devices is a major concern hampering their practical use in information technology.
Conductive metal oxides, such as RuO$_2$ and ReO$_3$, have been regarded as promising candidate materials for MEMS/NEMS contacts due to their conductivity, hardness, and relatively chemically inert surfaces. 
However, recent experimental works demonstrate that trace amounts of polymer could still form on RuO$_2$ surfaces.
In this study, 
we demonstrate the mechanism of this class of unexpected tribopolymer formation by conducting density functional theory based computational compression experiments with benzene as the contamination gas.
First, mechanical force during compression changes the benzene molecules from slightly physisorbed to strongly chemisorbed. 
Further compression causes deformation and chemical linkage of the benzene molecules.
Finally, the two contacts detach, with one having a complex organic molecule attached and the other with a more reactive surface.
The complex organic molecule, which has an oxabicyclic segment, can be viewed as the rudiment of tribopolymer, and the more reactive surface can trigger the next adsorption--reaction--tribopolymer formation cycle.
Based on these results, we also predict tribopolymer formation rates by using transition--state theory and the second--order rate law.
This study deepens our understanding of tribopolymer formation (especially on metal oxides) and provides strategies for suppressing tribopolymerization.
 
\end{abstract}

\section{Introduction}
Nano-- and microelectromechanical system (NEMS and MEMS) switches are potential next--generation electronic computing devices that could offer improved computational efficiency,
due to their small scale, low power consumption, and high speed~\cite{Liu12p39, Spencer11p308, Loh12p283, Kim15p673, Yoon15p217,Yeon16p3142,Ye16p235438}. 
Different from complementary metal--oxide semiconductor (CMOS) transistors, 
the on and off states of NEMS and MEMS are modulated by the closing and opening of the contacts, which could bridge the source and drain terminals into a closed circuit. 
In such an operation, there is essentially no open--circuit leakage or energy dissipation. 
However, after some switching cycles, tribopolymer (a kind of polymer caused by pressing and rubbing contacts together mechanically) forms on the surfaces of the contacts and severely reduces their conductivity~\cite{Streller14p13000120}. 
This effect hampers the practical application of NEMS and MEMS, and therefore understanding the underlying mechanism of tribopolymer formation is an on--going research challenge.

Tribopolymerization has been primarily studied in the context of lubricant additives; 
theoretical models have been developed by using both classical and quantum molecular dynamics~\cite{Furey05p621,Minfray08p589,Onodera10p034009,Koyama06p17507}. 
Some of the essential steps in tribopolymerization involve chemical bond breaking and formation.
Therefore, first--principles based methods are a powerful tool with distinct advantages for understanding the tribopolymer formation mechanism. 
Our previous study based on density functional theory calculations~\cite{Qi16p7529} demonstrated that reactive metals (Pt) can adsorb contamination gas molecules onto its surface and anchor them tightly.
Mechanical load during contact closure along with the effect of the metal as a catalyst together trigger the polymerization reaction~\cite{Qi16p7529}. 
Generally, conductive metal oxides, especially those with oxygen terminations, have less reactive and catalytic surfaces. 
Contamination gas molecules are less likely to be adsorbed, let alone undergo reaction.
However, recent experimental study observed a trace amount of polymer formed on RuO$_2$ contacts, especially when the benzene concentration is high~\cite{Brand13p1248}. 
This indicates that tribopolymer formation on conductive oxide surfaces has a different and not well understood mechanism.
In this study, we perform DFT calculations modeling a mechanical switching cycle, which involves a closure and detachment of the contacts, and propose  reaction pathways for tribopolymer formation on RuO$_2$(110).
Moreover, based on our proposed reaction path and calculated activation energies, we provide an estimate of tribopolymer formation rates at different contaminant gas concentrations and stresses. 

\section{Computational Methods }
Density functional theory (DFT) calculations are performed with the Q\textsc{uantum-espresso} code~\cite{Giannozzi09p395502}. 
Norm-conserving, optimized~\cite{Rappe90p1227}, designed nonlocal~\cite{Ramer99p12471} pseudopotentials are constructed with the OPIUM code~\cite{Opium} for all the elements in the system.
The exchange-correlation energy of electrons is included via the generalized gradient approximation (GGA) density functional of Perdew, Burke, and Ernzerhof~\cite{Perdew96p3865}. 
The kinetic energy cutoff is 680 eV. 
The self--consistent relaxation calculations are converged to a total force threshold of 5 meV/\AA. 
For the bulk structure, an $8\times8\times8$ Monkhorst-Pack mesh of $k$--points is used.
The optimized lattice parameters of RuO$_2$ ($P42/mnm$) are $a$=$b$=4.4681 \AA, $c$=3.0832 \AA\  and $\alpha$=90.0$^{\circ}$; 
the lattice parameters agree with the experimental values $a$=$b$=4.4919 \AA, $c$=3.1066 \AA\ and $\alpha$=90.0$^{\circ}$ ~\cite{Boman70p116} with an error of 0.53$\%$ in the $a$, $b$--directions and 0.75$\%$ in the $c$--direction. 

\begin{figure}[htpb]
\caption{Supercells for model surface and compression analysis. 
(a) Ball--and--stick model of the RuO$_2$(110)--O$^{\text{cvd}}$ surface with a full oxygen coverage. The superscripts `ot', `3f' and `brg' refer to on--the--top, three--fold coordinated, and bridging oxygens respectively. 
(b) Schematic representation of our computational compression model.
The initial supercell is 30 \AA\ long with a 20 \AA\ vacuum space. The three O--Ru bilayers slab contains 11 atomic layers, 
judging by the location of atoms along the $z$--direction.  
To simulate compression, the length of the supercell is reduced by $\Delta a$, with each inter--layer distance reduced by $\frac{\Delta a}{n-1}$, 
where $n$ is the number of atomic layers. The length of the centre vacuum region does not change before relaxation. Oxygen atoms are in red and Ru atoms are in light grey, whereas C atoms are in brown and H atoms are in pink.}
\begin{center}
\includegraphics[angle=0,width=1.0\columnwidth]{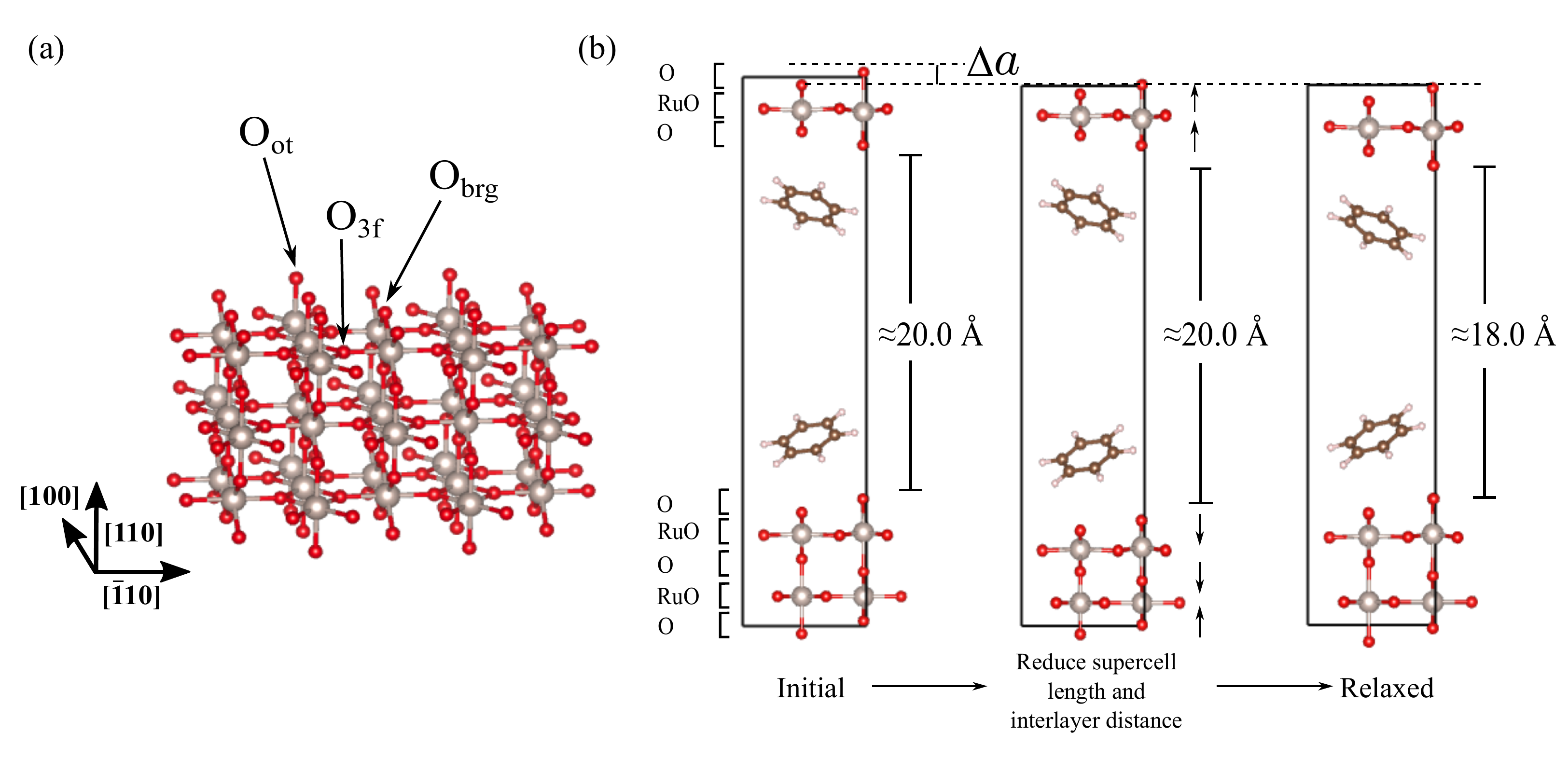}
\label{f1}
\end{center}
\end{figure}

For the surface model of the metal oxides, we use symmetric slabs for the RuO$_2$(110)-O$^{\text{cvd}}$ surface, as shown in Fig.~\ref{f1}, 
since such a surface is the most stable case under experimental conditions~\cite{Reuter02p035406,Kim17p1585}. The `cvd' in RuO$_2$(110)-O$^{\text{cvd}}$ refers to fully oxygen-``covered" the surface. 
The surface structure contains three rutile (RuO)-O bilayers capped with O. 
Benzene is chosen as the background contamination gas, since it comes from NEMS packaging and causes the most severe contamination~\cite{Hermance58p739}.
The surface is fully relaxed with a 20 \AA\ vacuum space between slabs, and a $8\times8\times1$ $k$--point grid is used to integrate the Brillouin zone.

In order to apply normal stress in simulations, we reduce the supercell height and the inter--layer distances of the metal oxide slab (see Fig.~\ref{f1}). 
After relaxation,  the strong Ru--O bonds tend to recover to their original inter--layer distances. 
As a result, the contamination gas molecules in the vacuum space sense the mechanical load along the $z$--direction.

After acquiring the reaction path from the compression computational experiments, the activation energy is computed with the nudged elastic band (NEB) method, to estimate the polymerization rates~\cite{Henkelman01p9657, Henkelman00p9901}.

\section{Computational Mechanical Cycle}

\begin{figure}[htpb]
\caption{
Schematic picture demonstrating a mechanical cycling,
which involves a compression and an expansion process. 
The upper panel shows the full--overlap case, in which the upper and lower benzene molecules are mirror images of each other.
The lower panel shows the non--overlap case, in which there is no overlap between the two benzene molecules. 
On the left of both upper and lower panels,  2$\times$2$\times$1 supercell is shown for describing two initial benzene adsorption registries on the RuO$_2$--O$^{\text{cvd}}$ surface. 
The supercell length is changed between 30.0 \AA\ and 14.0 \AA. 
Compression and expansion have the same step size change.
In each step, the change of the supercell length is 2 \AA\ between $c=30.0$ \AA\ and $c=18.0$ \AA. 
This change is reduced to 0.2 \AA\ for $15.0<c<16.0$ \AA,
and further reduced to 0.1 \AA\ when $c<15.0$ \AA, in order to have a more detailed observation of the reaction. Note that the RuO$_2$ surface slab is shown in stick style, with red representing O atoms and grey for Ru atoms.}
\begin{center}
\includegraphics[angle=0,width=1.0\columnwidth]{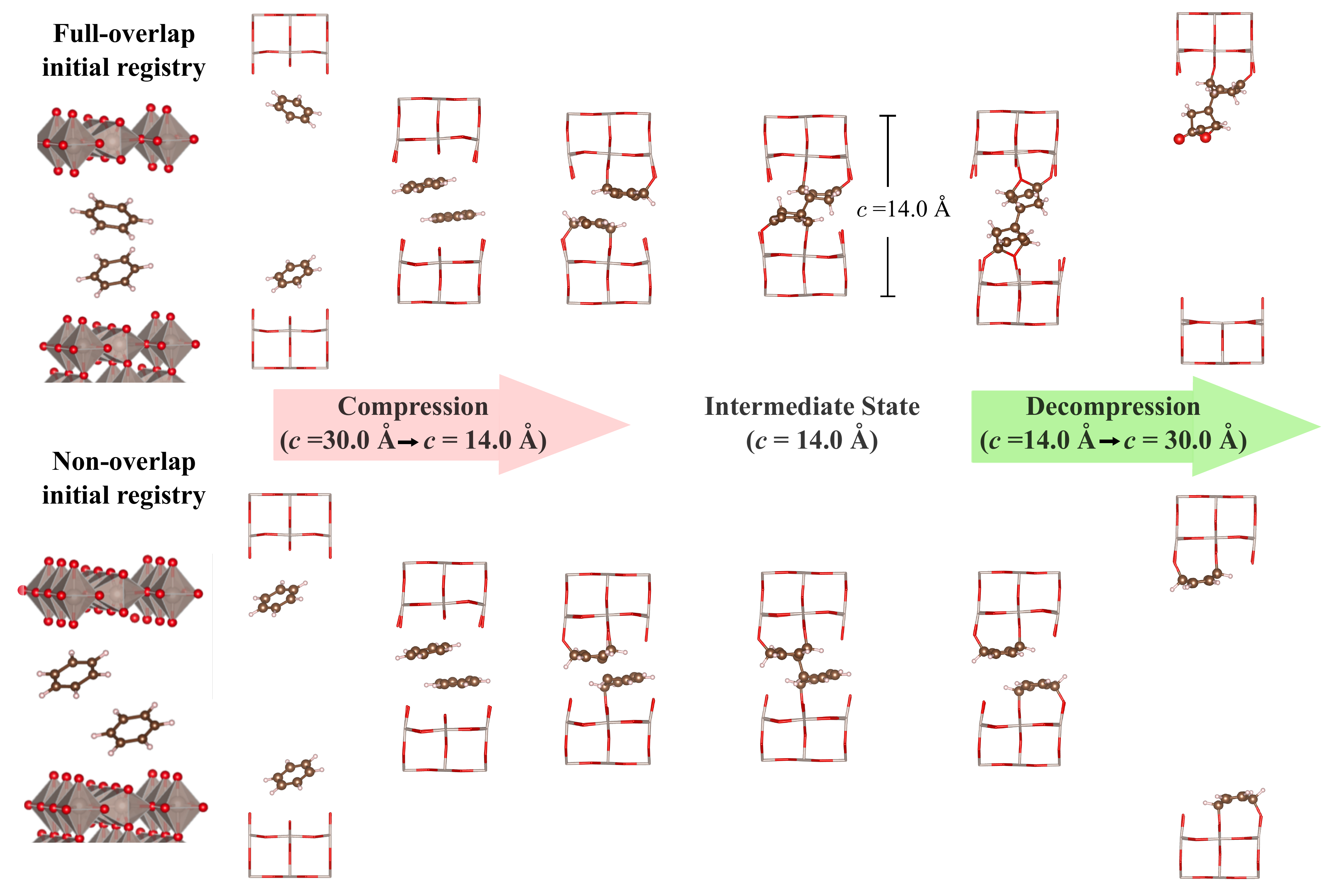}
\end{center}
\label{fig:schem}
\end{figure}

To search all the possible reaction paths, we started with two different initial benzene registries, which are referred to as full--overlap and non--overlap respectively, as shown in Fig.~\ref{fig:schem}. 
Computational mechanical switching cycles for both registries is shown in Fig.~\ref{fig:schem}. 
During the compression from $c=30$ \AA\ to $c=14$ \AA, benzene molecules in the two registries behave approximately the same.
Upon compression from $c=30$ \AA\ to $c=14.6$ \AA, the molecules rotate and slide due to the inter--molecular repulsion, and finally become approximately parallel to the surface. 
Upon further compression ($c=14.6$ \AA\ to $c=14.0$ \AA), the originally physisorbed benzene molecules begin to chemically bond with surface oxygen atoms.

\begin{figure}[htpb]
\caption{
Ball--and--stick model showing different intermediate--state structures. 
For clarity, we use thin lines to represent the RuO$_2$ substrate. 
(a) Intermediate state with H migrations. 
The green arrows show the hydrogen atom migrations between bottom and upper benzenes. 
(b) Intermediate state without H migrations. 
Only C--C bond forms between the upper and lower benzenes, as shown in the blue circle. 
(c) the intermediate state molecule (left) and its structural change (right) during a mechanical cycling. 
The carbon atom connecting the upper and lower benzene is labeled as $\alpha$-C. 
The ortho-C attaches to O$^{\text{brg}}$, and the meta-C attaches to O$^{\text{ot}}$.
}
\begin{center}
\includegraphics[angle=0,width=1.0\columnwidth]{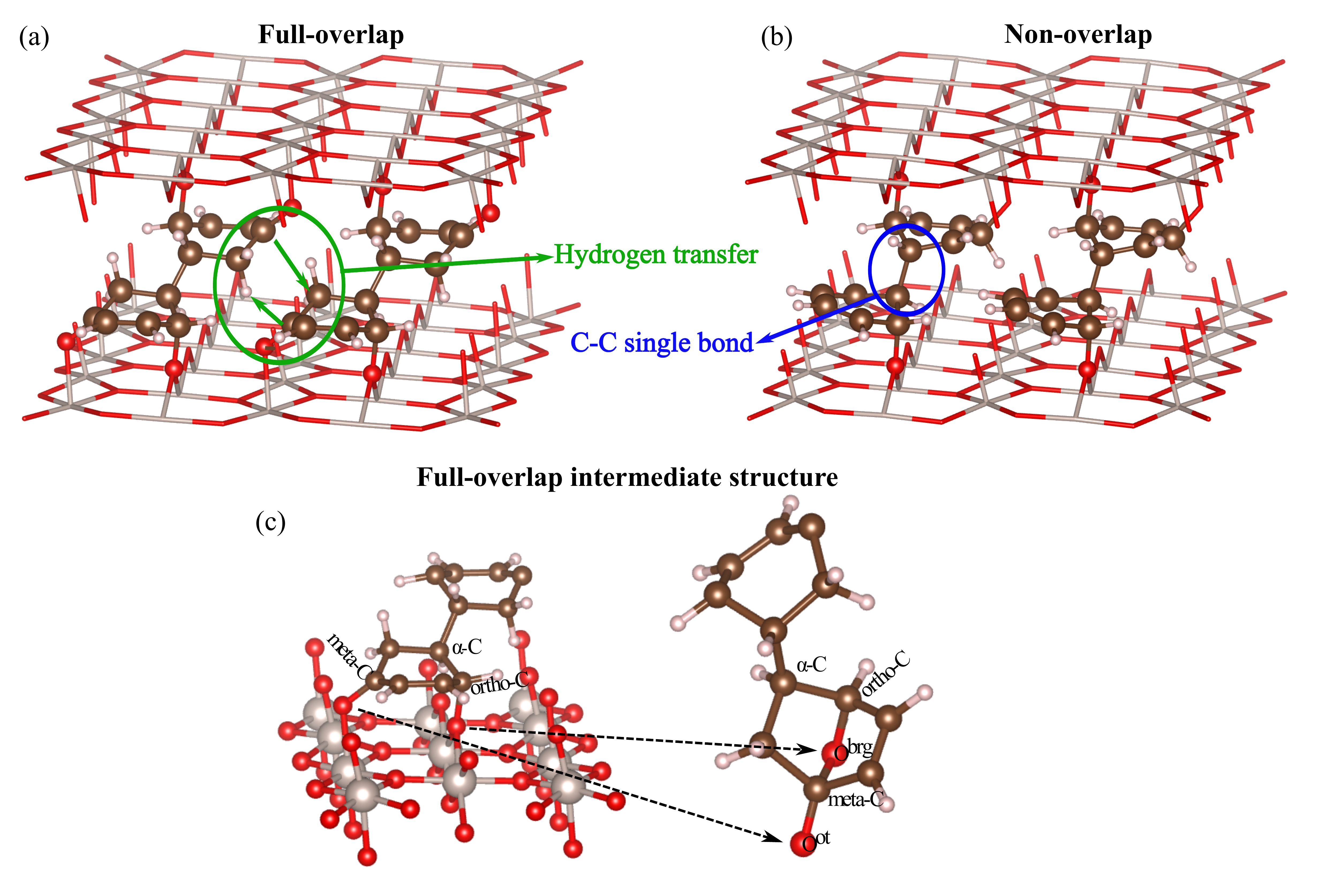}
\end{center}
\label{fig:inter}
\end{figure}

As the cell length is further compressed below 14.6 \AA, the two benzene molecules dimerize by forming a C--C bond.
The structure of this intermediate state formed by the linkage of two benzene molecules varies with the initial registry.
For a compression cycle starting with a full--overlap registry, 
there are hydrogen atom migrations between the bottom and the top benzene molecules, 
as shown in Fig.~\ref{fig:inter} (a). 
For the non--overlap initial registry compression, no such H atom migrations are observed during the intermediate state formation, as shown in Fig.~\ref{fig:inter} (b).

During the expansion ($c=14$ \AA\ to $c=30$ \AA ), we found that the intermediate biphenyl--like structure will either split into two adsorbed benzene molecules or become a biphenyl--like structure with an oxanorbornadiene segment, depending on its structure.
For the intermediate state with H migrations (whose structural evolution and the denomination of atoms are shown in detail in Fig.~\ref{fig:inter} (c)), 
the two O atoms which bond with the ortho--C and meta--C are extracted from the substrate. 
The O$^{\text{brg}}$ atom attached to the ortho--C originally also bonds to the meta--C atom, transforming the lower benzene ring into an oxanorbornadiene segment. 
This oxidized biphenyl--like molecule demonstrates that benzene molecules react under stress to form a complex polymeric precursor, 
which is the rudiment of tribopolymer.
Tribopolymer formation, therefore, involves a chain reaction of this molecule linkage process. 
H migration, during the formation of the intermediate state, plays an important role in inducing O extraction reaction. 
After H atoms migrate away from the ortho--C and meta--C, these two carbon atoms share more electrons and form stronger chemical bonds with the surface O atoms, 
thereby weakening the chemical bonds between O atoms and the RuO$_2$ substrate.
As a result, during expansion, two O atoms will be extracted from the surface and an oxanorbornadiene segment forms.
For an intermediate state without H migrations, the biphenyl--like structure splits into two strongly chemisorbed benzene molecules,
which could also lead to further polymerization as more cycles are applied.
Since the polymerization process for strongly chemisorbed molecules is well studied in our previous work~\cite{Qi16p7529},
we will not discuss it further.  

\section{Chemical Kinetics}

We will now focus on the case in which an oxidized biphenyl--like molecule is produced, 
and estimate the time--scale of tribopolymerization. 
This reaction involves two steps. First, the benzene molecules change from weakly physisorbed gas moelcules, 
which interact with the contacts by weak Van der Waals interaction, to strongly--chemisorbed adsorbates,
\begin{equation}
\text{C}_{\text{6}}\text{H}_{\text{6(g)}}\rightarrow \text{C}_{\text{6}}\text{H}_{\text{6(ads)}},
\label{eq_ads}
\end{equation}
and then two chemically adsorbed benzene molecules follow a second--order reaction as,
\begin{equation}
\text{2C}_{\text{6}}\text{H}_{\text{6(ads)}}\rightarrow \text{C}_{\text{12}}\text{H}_{\text{12(ads)}}.
\label{eq_rxn}
\end{equation}
Gas chemisorption depends on the partial pressure and the chemisorption energy. 
We consider the simplest relationship, the Langmuir isotherm, between benzene molecule surface coverage, $\theta$, and to benzene pressure, $p_{\text{C}_6\text{H}_6}$, as follows:
\begin{equation}
\theta_0=\frac{p_{\text{C}_\text{6}\text{H}_{\text{6}}}}{p_{\text{C}_\text{6}\text{H}_{\text{6}}}+p_{\text{0}}},
\end{equation}
where $\theta_0$ represents the initial equilibrium coverage, $p_{\text{C}_\text{6}\text{H}_{\text{6}}}$ is the initial partial pressure of benzene in the gas phase and $p_{\text{0}}$ takes the form as,
\begin{equation}
p_{\text{0}}=\left(\frac{2\pi mk_{\text{B}}T}{h^2}\right)^{\frac{1}{2}}k_{\text{B}}T\exp\left(-\frac{E_{\text{ads}}}{k_{\text{B}}T}\right).
\end{equation}
Here, $m$ is the molecular weight of benzene, $k_{\text{B}}$ is the Boltzmann constant, $T$ is the temperature, 
$h$ is Planck's constant, and $E_{\text{ads}}$ is the adsorption energy of benzene on the surface. 
The rate constant, $k$, for the intermediate state formation (Eq.~\ref{eq_rxn}), treated by transition state theory, can be written as,  
\begin{equation}
k=\frac{k_{\text{B}}T}{h}\exp\left(-\frac{\Delta E(\sigma)}{k_{\text{B}}T}\right),
\label{eq6}
\end{equation}
where $\Delta E$ is the activation energy under applied normal stress, $\sigma$. Since Eq.~\ref{eq_rxn} is a second-order reaction, the rate law can be written as
\begin{equation}
\text{rate}=-\frac{d\theta(t)}{dt}=2k\theta(t)^2,
\end{equation}
where $\theta$ is the surface coverage of adsorbed benzene and $k$ is the rate constant. 

Here, we introduce the reaction half--life time ($t_{1/2}$), 
which is defined as the time that half of the surface adsorbates are consumed for tribopolymerization. 
$t_{1/2}$ is expressed as
\begin{equation}
t_{\frac{1}{2}}=\frac{1}{2k\theta_0}.
\label{eq:half_life}
\end{equation}

\begin{figure}[ht]
\caption{(a) The activation energy (red dots) and normal stress (blue squares) {\em{vs.}} supercell length. The purple dashed line is the linear fit of the activation energy.
(b) Calculated normal-stress-dependent reaction rates. The threshold stress for a zero--barrier reaction change is 15 GPa. The blue dashed lines are the linear fits.
(c) Reaction half--life under different benzene partial pressures and normal stresses. }
\begin{center}
\includegraphics[angle=0,width=0.85\columnwidth]{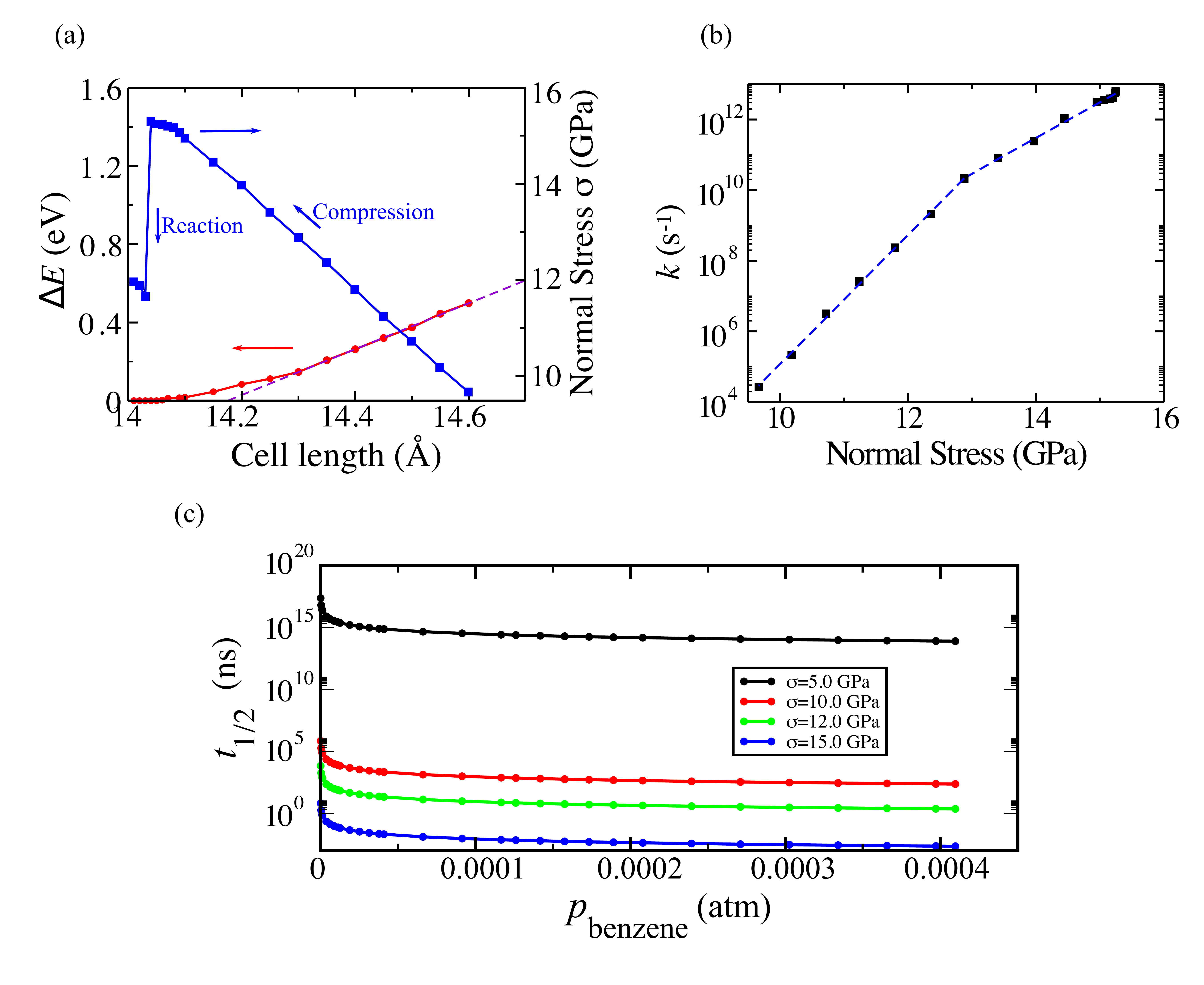}
\end{center}
\label{fig:stress_rconst}
\end{figure}

To acquire $k$, we calculate $\Delta E(\sigma)$ at different supercell lengths with the NEB method, as shown  
in Fig.~\ref{fig:stress_rconst} (a).
We observe that the activation energy is lower for a smaller supercell.
Normal stress {\em{vs.}} supercell length is also plotted. During the compression, the normal stress increases linearly and plateaus around 15.0 GPa. 
Further compression causes the normal stress drop to 11.66 GPa, indicating that the molecule--linkage reaction occurs.
This stress, 15.0 GPa, is the threshold for a zero--barrier reaction, above which polymerization reaction will occur spontaneously.
This value (15.0 GPa) is also consistent with the estimated stress in experiments~\cite{Zhu11p1914,Brand13p341}. 

By using Eq.~\ref{eq6} and Eq.~\ref{eq:half_life}, we calculated reaction rate constant $k$ and corresponding reaction half--life at different benzene partial pressures and applied normal stresses, 
as shown in Fig.~\ref{fig:stress_rconst} (b) and Fig.~\ref{fig:stress_rconst} (c).
Smaller applied stresses and contamination gas molecule partial pressures lead to slower tribopolymer formation.
The half--life decreases exponentially with benzene partial pressure, and plateaus at and above $1\times10^{-4}$ atm, indicating the saturation of surface adsorption sites.
Normal stress, which could affect the half--life by several orders of magnitude, is more influential than partial pressure.
Therefore, 
techniques for reducing normal mechanical load, such as controlling the momentum of NEMS contacts and fabricating flatter contact surfaces (high spots on the surfaces may cause the stress to peak there) will be of great significance.

\section{Conclusion}
We carry out DFT calculations to reveal the tribopolymer formation mechanism on the RuO$_2$(110)-O$^{\text{cvd}}$ surface. 
We find that normal stress changes molecular adsorption from weak physisorption to strong chemisorption, 
and sufficient stress (15 GPa) will cause  benzene molecules link to form an intermediate biphenyl--like state,
which can adopt two different structures.
One, with H migration, will change to a biphenyl--like molecule with an oxabicyclic segment, which is the rudiment of tribopolymer.
Another structure, without H migration will split into two adsorbed benzene molecules.
Further, we predict the reaction rates based on the transition--state theory and second--order rate law.
This study provides more insights about the mechanochemical reactions of tribopolymerization on conductive oxides.

\section{Acknowledgements}
J.Y. was supported by the U.S. National Science Foundation, under grant CMMI-1334241.
H.D.K. was supported by the University of Pennsylvania NSF/Louis Stokes Alliance for Minority Participation (LSAMP) program.
Y.Q. was supported by the U.S. National Science Foundation, under grant DMR-1124696.
A.M.R. was supported by the U.S. Department of Energy, under grant DE-FG02-07ER15920.
Computational support was provided by the HPCMO of the U.S. DOD and the NERSC of the U.S. DOE.

\bibliography{acs-ruo2-compression}

\end{document}